# The worldwide NORM production and a fully automated gamma-ray spectrometer for their characterization

Xhixha G. [1,d,e,f], Bezzon G. P.[b], Broggini C.[a], Buso G. P.[b], Caciolli A.[a,c], Callegari I.[c], De Bianchi S. [d], Fiorentini G. [b,d,e], Guastaldi E.[c], Kaçeli Xhixha M.[g], Mantovani F.[d,e], Massa G.[c], Menegazzo R.[a], Mou L.[c], Pasquini A.[c], Rossi Alvarez C.[a], Shyti M.[d,e]

[a] Istituto Nazionale di Fisica Nucleare (INFN), Padova Section, Via Marzolo 8 - 35131 Padova, Italy.
[b] Istituto Nazionale di Fisica Nucleare (INFN), Legnaro National Laboratory, Via dell'Università, 2 - 35020 Legnaro, Padova, Italy.
[c] University of Siena, Center for GeoTechnologies, Via Vetri Vecchi, 34 - 52027 San Giovanni Valdarno, Arezzo, Italy.
[d] University of Ferrara, Physics Department, Via Saragat, 1 - 44100 Ferrara, Italy.
[e] Istituto Nazionale di Fisica Nucleare (INFN), Ferrara Section, Via Saragat, 1 - 44100 Ferrara, Italy.
[f] Agricultural University of Tirana, Faculty of Forestry Science, Kodër Kamëz - 1029 Tirana, Albania.
[g] University of Sassari, Botanical, Ecological and Geological Sciences Department, Piazza Università 21- 07100 Sassari, Italy.

**Abstract**

Materials containing radionuclides of natural origin and being subject to regulation because of their radioactivity are known as NORM (Naturally Occurring Radioactive Material). By following IAEA, we include in NORM those materials with an activity concentration, which is modified by human made processes. We present a brief review of the main categories of non-nuclear industries together with the levels of activity concentration in feed raw materials, products and waste, including mechanisms of radioisotope enrichments. The global management of NORM shows a high level of complexity, mainly due to different degrees of radioactivity enhancement and the huge amount of worldwide waste production. The future tendency of guidelines concerning environmental protection will require both a systematic monitoring based on the ever-increasing sampling and high performance of gamma ray spectroscopy. On the ground of these requirements a new low background fully automated high-resolution gamma-ray spectrometer MCA_Rad has been developed. The design of lead and cooper shielding allowed to reach a background reduction of two order of magnitude with respect to laboratory radioactivity. A severe lowering of manpower cost is obtained through a fully automation system, which enables up to 24 samples to be measured without any human attendance. Two coupled HPGe detectors increase the detection efficiency, performing accurate measurements on small sample volume (180 cc) with a reduction of sample transport cost of material. Details of the instrument calibration method are presented. MCA_Rad system can measure in less than one hour a typical NORM sample enriched in U and Th with some hundreds of Bq/kg, with an overall uncertainty less than 5%. Quality control of this method has been tested. Measurements of three certified reference materials RGK-1, RGU-2 and RGTh-1 containing concentrations of potassium, uranium and thorium comparable to NORM have been performed. As a result, this test achieved an overall relative discrepancy of 5% among central values within the reported uncertainty.

## 1. Introduction

Materials containing radionuclides of natural origin and being subject to regulation because of their radioactivity are known as NORM (Naturally Occurring Radioactive Material).

---

[1] Xhixha Gerti, Dipartimento di Fisica, Università di Ferrara, Polo Scientifico e Tecnologico Via Saragat, 1 – 44100 Ferrara, Italy. Phone: +39-3200864636, Fax: +39-0532974210, E-mail: xhixha@fe.infn.it

By following IAEA, we include in NORM those materials with an activity concentration altered by human made processes[2] (**IAEA, 2008; IAEA, 2003a**). In the last decades the large production of NORM and the potential long-term radiological hazards, due to long-lived radionuclides, represented an increasing level of concern. The development of instruments devoted to the measurements of NORM concentrations is a crucial task for the evaluation of the radiological impact on both workers and public members.

NORM are found as products, by-products and/or wastes of industrial activities, such as production of non-nuclear fuels (e.g. coal, oil and gas), mining and milling of metalliferous and nonmetalliferous ores (e.g. aluminum, iron, cooper, gold and mineral sand), industrial minerals (e.g. phosphate and clays), radioisotope extraction and processing, as well as water treatments (**IAEA, 2003b**).

The most important sources of natural radioactivity are due to the presence of $^{238}$U, $^{232}$Th and $^{40}$K in the Earth. Generally $^{235}$U and $^{87}$Rb and other trace elements are negligible. The decay chain of $^{238}$U ($^{232}$Th) includes 8 (6) alpha decays and 6 (4) beta decays respectively, which are often associated with gamma transitions. The detection of these radioactivity sources can be performed through a wide set of methods, such as gamma-ray spectroscopy, alpha spectroscopy, neutron activation analysis (NAA), inductively-couple plasma mass-spectroscopy (ICP-MS), inductively coupled plasma atomic emission spectroscopy (ICP-AES), X-ray fluorescence spectroscopy (XRF) and liquid scintillation counting (LSC) (**IAEA, 2006**). The choice of the methodology for determining radioactive content of NORM depends on many factors, especially the economic character and prompt measurement of the individual samples.

Usually, in NORM $^{238}$U and $^{232}$Th decay chains are not in secular equilibrium. It means that in $^{238}$U decay chain, some long lived radionuclides ($^{238}$U, $^{226}$Ra, $^{210}$Pb) represent the head of decay chain segments, which can reach the secular equilibrium in less than one year: the gamma ray spectrometry, therefore, is the suitable technique for measuring the abundance of these radionuclides and for checking the secular equilibrium among the respective chain segments. NORM can be enriched in $^{230}$Th and $^{210}$Po, which can be out of chain segments: the activities of $^{230}$Th can be determined by gamma-spectrometry directly, while the suitable technique for quantifying $^{210}$Po concentration is alpha spectrometry. In $^{232}$Th decay chain the two chain segments, having on head $^{228}$Ra and $^{228}$Th, reach the secular equilibrium in about one month. By measuring the gamma transitions in each chain segment, one can determine $^{228}$Ra and $^{228}$Th abundances; while XRF, NAA, ICP-AES, ICP-MS are suitable techniques for measuring $^{232}$Th content.

By using high-resolution gamma-ray spectrometry, all radioisotopes of $^{238}$U and $^{232}$Th decay chains, with the exception of $^{232}$Th and $^{210}$Po, can be investigated simultaneously. The radioactivity characterization of NORM may require a massive amount of measures of samples. For this purpose we designed and built up a low background high-resolution gamma-ray spectrometry system, which allows an autonomous investigation of the radioactivity content on a large amount of samples, without any human attendance.

An empirical method for the characterization of the absolute efficiency of this instrument is presented in detail. Along with the instrument calibration, a quality test of this method was carried out. We tested the performances of the instrument by using three reference materials RGK-1, RGU-1 and RGTh-1 certified by International Atomic Energy Agency (IAEA) and containing radioactive concentration comparable to NORM values.

## 2. Industrial processes producing NORM: an overview

---

[2] Sometimes these materials are known in the literature as TENORM (technologically enhanced naturally occurring radioactive material).

## 2.1 Non-nuclear fuels extraction and processing

*Oil and gas industry*

Most hydrocarbons are trapped within porous reservoirs (known as oil/tar sand and oil shale deposits) by impermeable rocks above: the rock formations holding the oil also contain U and Th at the order of some ppm, corresponding to a total specific activity of some tens[3] of Bq kg$^{-1}$. Oil and gas reservoirs contain a natural water layer (formation water) that lies under the hydrocarbons: U and Th do not go into solution, but the formation water tends to reach a specific activity of the same order of the rock matrix (**Metz et al., 2003**) due to dissolution of $^{226}$Ra and $^{228}$Ra radium as radium chloride (**Heaton and Lambley, 1995**). Additional heated water is often injected into the reservoirs to achieve maximum oil recovery. This process disturbs the cation/anion ratio and alters the solubility of various sulfate salts, particularly $BaSO_4$ and $RaSO_4$: radium co-precipitates with barium as sulfates forming scale within the oil pipes (**Jerez Vegueria et al., 2002**). The solids, which are dissolved in crude oil and in the produced water, precipitate forming the sludge, which is a mixture of oil, carbonates and silicates sediments, as well as corrosion products that accumulate inside piping and in the bottom of storage tanks. The contribution to the radioactivity content in the sludge comes mainly from the precipitates of hard insoluble radium sulphate and possibly from radioactive silts and clays (**Heaton and Lambley, 1995**). The specific activity of scales and sludges can vary enormously and generally it is of several orders of magnitude more than formation water. As reported in **Table 1,** the high variability, often among the samples collected in the same area, shows that a frequent and dense sampling of these NORM may be required.

*Coal-fired power plant*

Coal is a combustible sedimentary rock formed through the anaerobic process of the decomposed dead plants accumulated at the bottom of basins of some marsh land, lake or sea: the coalification process yields a product rich in carbon and hydrogen (**Rubio Montero et al., 2009**). The organic matter plays an important role in the uranium concentrations at syngenetic, epigenetic and diagenetic stages of the sedimentary cycle (**Nakashima, 1992**). The complexation and reduction of uranium are considered the main geochemical processes of the fixation of uranium on organic matter from very dilute solutions. The complexation of uranium involves the uranyl cation ($UO_2^{2+}$) producing a $UO_2$ complex through a dehydrogenation process of organic matter. During the reduction of the soluble $UO_2^{2+}$, the insoluble $UO_2$ is produced precipitating as uraninite (**Landais, 1996**). The efficiency of these processes depends on the chemistry of the organic matter and on the temperature of reaction, producing U concentrations ranged over a couple orders of magnitude: the typical range of radionuclide activity concentrations of $^{238}$U, $^{232}$Th and $^{40}$K in coal can be 10-600 Bq kg$^{-1}$, 10-200 Bq kg$^{-1}$ and 30-100 Bq kg$^{-1}$ respectively (**Beck, 1989**). The geological processes over the time increase the grade of coal transforming the organic material from peat to graphite. In low-grade coal the secular equilibrium between $^{238}$U and $^{232}$Th and their decay products is not expected while in the high-grade coal it may exist (**Tadmor, 1986**). The volatilization–condensation process of particles during coal combustion breaks the secular equilibrium and increases the radionuclide concentrations with decreasing of particle size: the maximum enrichment has been measured in particles with diameters of about 1 μm. $^{210}$Pb and $^{210}$Po exhibit the greatest enrichment, as much as a factor of 5, while maximum enrichment for uranium isotopes is about a factor of 2, and for $^{226}$Ra a factor of around 1.5. In some samples of fly ash the $^{210}$Pb and $^{226}$Ra activity exceeds thousands of Bq kg$^{-1}$ (**Flues, 2006**).

---

[3] The U and Th abundances vary for different rock types: e.g. in considering shale, the range of U and Th abundances reported by (**Condie, 1993**) are 2.4-3.4 ppm and 8.5-14.3 ppm respectively. We remind that the specific activity of 1 ppm U (Th) corresponds to 12.35 (4.06) Bq/kg.

According to the World Coal Institute (**WCI, 2005**) 40% of the world's electricity in 2003 is generated by coal: in the same year the world coal consumption reached $48.4 \cdot 10^{11}$ kg (**U.S. EIA / IEO, 2010**). After the combustion of the bituminous coal containing an average ash of 12%, coal by-products are composed by some 70% of fly ash and by some 30% of the bottom ash and boiler slag. A large fraction (more than 95%) of these small particles can be removed from gas stream, by usually applying electrostatic precipitator and fabric filters: in 2003 the estimated worldwide fly ash production was $3.9 \cdot 10^{11}$ kg (**Mukherjee et al., 2008**). In 2003 the US and EU fly ash production was about $0.6 \cdot 10^{11}$ kg and $0.4 \cdot 10^{11}$ kg respectively. In US and UE about 39% and 48% of this amount was recycled respectively (**ACAA, 2003**; **ECOBA, 2003**). According to (**U.S. EIA / IEO, 2010**) the world coal consumption in 2035 is projected to reach $93.8 \cdot 10^{11}$ kg, corresponding to an estimated fly ash production of $7.5 \cdot 10^{11}$ kg. This large amount of NORM is required to be measured and monitored with accuracy, by developing fast and *ad hoc* methods: the MCA_Rad system presented in section 3 has been designed as a response to this increasing demand.

**Table 1**. Range of specific activity concentrations of $^{226}$Ra and $^{228}$Ra in scale and sludge, as reported by different authors for various geographic regions. Remind that the exemption level recommended by (**IAEA, 1996**) is $10^4$ Bq kg$^{-1}$ for both $^{226}$Ra and $^{228}$Ra.

| Authors of the study | Country | $^{226}$Ra [$10^4$ Bq kg$^{-1}$] | | $^{228}$Ra [$10^4$ Bq kg$^{-1}$] | |
|---|---|---|---|---|---|
| | | Scale | Sludge | Scale | Sludge |
| **Godoy and Cruz, 2003** | Brazil | 1.91 – 32.3 | 0.036 – 36.7 | 0.421 – 23.5 | 0.025 – 34.3 |
| **Gazineu et al., 2005a** | Brazil | 12.1 – 95.5 | 0.24 – 350 | 13.1 – 79.2 | 205 |
| **Gazineu and Hazin, 2008** | Brazil | 7.79 – 211 | 0.81 – 41.3 | 10.2 – 155 | 0.94 – 11.8 |
| **Shawky et al., 2001** | Egypt | 0.754 – 14.3 | 0.0018 | 3.55 – 66.1 | 1.33 |
| **Abo-Elmagd et al., 2010** | Egypt | 49.3 – 51.9 | 0.527 – 0.886 | 3.20 – 5 | 0.1 – 0.19 |
| **Bakr, 2010** | Egypt | 0.0016 – 0.0315 | 0.00055 – 0.179 | 0.00007 – 0.0177 | 0.00007 – 0.0885 |
| **Omar et al., 2004** | Malaysia | 0.055 – 43.4 | 0.0006 – 0.056 | 0.09 – 47.9 | 0.0004 – 0.052 |
| **Lysebo et al., 1996** | Norway | 0.03 – 3.23 | 0.03 – 3.35 | 0.01 – 0.47 | 0.01 – 0.46 |
| **Al-Saleh and Al-Harshan, 2008** | Saudi Arabia | 0.00008 – 0.00015 | 0.00068 – 0.00594 | 0.000014 – 0.00031 | 0.00063 – 0.00476 |
| **Al-Masri and Suman, 2003** | Syria | 14.7 – 105 | 47 – 100 | 4.3 – 18.1 | 35.9 – 66 |
| **Jonkers et al., 1997** | USA | 0.01 – 1500 | 0.005 – 80 | 0.005 – 280 | 0.05 – 50 |
| **Zielinski et al., 2001** | USA | 1.88 – 489 | 2.04 – 6.38 | 0.118 – 19 | 0.241 – 0.574 |

## 2.2 Metal mineral extraction and processing

*Bauxite extraction and alumina production*

In metal mining and waste processing, the radioactive content varies from $10^{-2}$ kBq kg$^{-1}$, for the large volume industry, to $10^2$ kBq kg$^{-1}$ for rare earth metals (**IAEA, 2003b**). In massive metal extraction, the amount of NORM produced by bauxite processing is relevant.

Bauxites generally contain concentrations of Th and U greater than the Earth's crustal average: in a multi-methodological study published by (**Adams and Richardson, 1960**) based on twenty-nine samples of bauxites from different locations, the reported specific activities of U and Th are in the range 33 – 330 Bq kg$^{-1}$ and 20 – 532 Bq kg$^{-1}$ respectively. These values can be compared with the typical concentration in bauxite: 400 – 600 Bq kg$^{-1}$ for U and 300 – 400 Bq kg$^{-1}$ for Th (**UNSCEAR, 2000**).

The parent rocks affect the U and Th abundances in the bauxites: in particular the bauxites derived from acid igneous rocks show a concentration higher than those extracted from basic igneous rocks, whereas the bauxites mined from deposits of shales and carbonates rocks are characterized by intermediate concentrations. The process of lateritization during bauxite formation contributes to increase the ratio Th/U, which is generally more than 4 (**Adams and Richardson, 1960**).

Bayer process is used for refining bauxite to smelting grade alumina, the aluminium precursor: it involves the digestion of crushed bauxite in a concentrated sodium hydroxide solution at temperatures up to 270°C. Under these conditions, the majority of the species containing aluminium is dissolved in solution in the ore, while the insoluble residues are filtered making a solid waste called "red mud". The alumina is obtained by the hydroxide solution after the processes of precipitation and calcination (**Hind et al., 1999**). Finally, by using the Hall-Heroult electrolytic process, alumina is reduced to aluminium metal.

The fact that thorium and radium in an hydroxide medium are practically insoluble (**Somlai et al., 2008**) could disturb especially the secular equilibrium of the uranium chain, as it has been observed in some measurements (**Pontikes et al., 2006**). However, some "results indicate that the chemical processing of the bauxite ore has not significant consequences in the secular equilibrium of either the uranium or thorium series" (**Cooper et al., 1995**). In the process of extracting alumina from bauxite, over 70% of the thorium and radium are concentrated in the red mud (**Adams and Richardson, 1960**).

In 2009 the worldwide production of bauxite and alumina was $199 \cdot 10^9$ kg and $123 \cdot 10^9$ kg respectively. Considering that the worldwide ratio bauxite/alumina, averaged in the period 1968-2009, is 2.7 ± 0.1, we expect that 1.7 ± 0.1 kg of red mud is generated per kg of alumina (**USGS Data Series 140**). By assuming that at least 70% of the radioisotopes in bauxite accumulate in the red mud, the increasing factor of radioactivity content in the red mud varies in a range of 1.1 and 1.6. Using these estimated enrichments for bauxite we encompass a large portion of the radium and thorium activities of the red mud reported in literature (**Table 2**).

Based on recent statistics, more than $70 \cdot 10^9$ kg of red mud is discharged annually in the world: it constitutes the most important disposal problem of the aluminium industry. The mud is highly basic (pH > 10) and its storage on huge area can cause environmental pollution, soil basification, paludification, surface water and groundwater pollution as well as resource pollution. The safe treatment of this NORM is an increasing social problem. Moreover, a considerable attention has been given to additional uses of bauxite wastes: they include metallurgical extractions, building materials productions and the development of new ceramics and catalytic materials. Gamma-ray spectrometers are able to process thousands of measurements in order to perform environmental monitoring and to control the recycled by-products. The MCA_Rad system described in section 3 could be extremely helpful for processing these measurements by significantly diminishing manpower costs.

**Table 2.** Activity concentration in red mud reported in the literature: in a) the range indicated correspond to the maximum and minimum value measured for different samples and in b) the uncertainties of measurement results. The activities quoted are assumed equal to $^{228}$Ac in 1), equal to $^{228}$Th in 2) and equal to $^{238}$U in 3).

| Authors of the study | Country | Activities (Bq kg$^{-1}$) | |
|---|---|---|---|
| | | $^{226}$Ra | $^{232}$Th |
| **Papatheodorou et al., 2005** [a)] | Greece | 13 – 185 | 15 – 412 |
| **Philipsborn and Kuhnast 1992** [b) 1)] | Germany | 122 ± 18 | 183 ± 33 |
| **Pinnock, 1991** [a)] | Jamaica | 370 – 1047 | 328 – 350 |
| **Somlai et al., 2008** [a)] | Hungary | 225 – 568 | 219 – 392 |
| **Akinci and Artir, 2008** [b)] | Turkey | 210 ± 6 | 539 ± 18 |
| **Jobbágy et al., 2009** [a)] | Hungary | 102 – 700 | 87 – 545 |
| **Cooper et al., 1995** [b) 2)] | Australia | 310 ± 20 | 1350 ± 40 |
| **Turhan et al., 2011** [a)] | Turkey | 128 – 285 | 342 – 357 |
| **Pontikes et al., 2006** [b) 2)] | Greece | 379 ± 43 | 472 ± 23 |
| **Beretka and Mathew, 1985** | Australia | 326 | 1129 |
| **Georgescu et al., 2004** | Romania | 212 | 248 |
| **Döring et al., 2007** [b)] | Germany | 190 ± 30 | 380 ± 50 |
| **Ruyters et al., 2011** [3)] | Belgium | 550 | 640 |

*Mineral sand and downstream productions*

The extraction of mineral sand ore is important for the production of heavy minerals (with densities heavier than 2.8 g cm$^{-3}$) like titanium, tin and zirconium bearing minerals and rare earth elements[4] (REEs). The deposits of hard minerals, which do not undergo erosion and transport processes, mainly occur when they have been concentrated by marine, alluvial and/or wind processes. These placer deposits can be found also in vein deposits, mostly disseminated in alkaline intrusions in hard rocks. The radioactivity concentration of mineral sand can be of the order of a few hundreds Bq kg$^{-1}$, depending on the placer geology. In heavy minerals we can often find high content of radioactivity, sometimes of the order of hundreds kBq kg$^{-1}$.

The process involved in heavy mineral extraction includes two main phases of separation. The first phase separates the heavy mineral concentration, by using either dry operation or dredging of the slurried ore: this produces high amount of residues with a radionuclide concentration of the same order of mineral sand (**Paschoa, 2008**). During the second phase, the heavy mineral concentration is further separated mainly by combining dry magnetic and electrostatic processes. This allows the concentration of various minerals, such as titanium bearing minerals (ilmenite, leucoxene, rutile), zircon bearing minerals (zircon, baddeleyite) and REEs bearing minerals (monaxite, zenotime). The products of this second phase generally show a high content of radioisotopes.

Titanium bearing minerals are used mainly to produce $TiO_2$ pigment. The radioactivity concentration in these minerals varies in rutile (400 - 2900 Bq kg$^{-1}$ Th and 250 – 500 Bq kg$^{-1}$ U) and ilmenite (400 – 4100 Bq kg$^{-1}$ Th and 250 – 750 Bq kg$^{-1}$ U) (**IAEA_2003b**; **Timmmermans and van der Sten, 1996**; **McNulty, 2007**). The higher production efficiency is obtained by rutile: it is directly processed through the chloride route producing $TiO_2$ pigment/waste with a ratio of 5/1 (**USGS MCS, 1996**). Since the radioisotopes follow the liquid waste stream, the radioactivity concentration in the waste is very high due to the severe mass reduction.

---

[4] REEs contain 16 chemical elements, including those with atomic numbers 57 (lanthanum) through 71 (lutetium), as well as yttrium (atomic number 39), which has similar chemical properties.

Ilmenite requires a pre-processing in order to produce synthetic rutile: the product/waste ratio is 10/7. The synthetic rutile is further processed through the chloride route in order to produce $TiO_2$ pigment with a product/waste ratio of 5/6, showing a light enhancement of radioactivity concentration in waste of ilmenite processing (**IAEA_2003b**).

In $TiO_2$ pigment production the most relevant NORM are made by the chloride treatment of rutile. The worldwide production of $TiO_2$ in 2010 was about $5.7 \cdot 10^9$ kg when the mineral extraction ratio is 1/10 for rutile/ilmenite (**USGS MCS, 2011; IAEA_2003b**): about 2% of the total waste which has been generated can show a strong enhancement in radioactivity concentration.

The most common zirconium bearing minerals are zircon and baddeleyite: in 2010 the world extraction of these minerals was about $1.2 \cdot 10^9$ kg (**USGS MCS, 2011**). Mineral sands containing zircon are commonly used in ceramic and refractory industries. The zircon crystal lattice host uranium and thorium: their activities in zircon bearing minerals are in the range of 1-5 kBq kg$^{-1}$ for U and of 0.5-1 kBq kg$^{-1}$ for Th respectively (**Righi et al., 2005**).

The manufacture of zirconia is mainly performed by fusion of feedstock with coke near to zirconium molten temperatures. It causes the dissociation of mineral in $ZrO_2$ and $SiO_2$: during the fusion U and Th end up almost at the same concentration in zircon product, while $^{226}$Ra tends to end up in silica, causing the disequilibrium of U decay chain. The subsequent caustic fusion process at 600°C increases the purity of zirconia. During the chemical dissolution, zircon crystal structure is destroyed, yielding nearly 100% of uranium recovery in the form of sodium uranate (**Brown and Costa, 1972**). The performances of MCA_Rad system completely fit the need of measuring radioactivity content in this kind of NORM.

The main minerals used as sources of REEs can be extracted by placer deposits (monazite and xenotime) and by hard rocks (bastnaesite, coperite and pyrochlore). Monazite minerals are characterized by high activity concentration of U (25-75 kBq kg$^{-1}$) and Th (41-575 kBq kg$^{-1}$). The chemical attack of the mineral based on NaOH separates sodium phosphate from a mixture called "cake I", which is rich in heavy minerals. Cake I is further filtered, given that it yields a concentration of rare earth chlorides and a mixture called "cake II" containing most of thorium and uranium originally present in monazite feedstock (**Paschoa, 1997**). The decay chain of U and Th are not in secular equilibrium in cake II: thorium precipitates while radium remains in solution. The latter can reach very high activities (7 MBq kg$^{-1}$ and 10 MBq kg$^{-1}$ for $^{226}$Ra and $^{228}$Ra respectively) (**Paschoa, 2008**). In 2010, the extraction of monazite has been $1.3 \cdot 10^8$ kg (**USGS MCS, 2011**) and its processing has made 10% of cake II waste (**Paschoa, 1993**). This large amount of highly enhanced radioactivity NORM requires scrupulous monitoring and control.

*2.3 Industrial minerals extraction and processing*

*Phosphate fertilizer industry*

The main phosphate-rock (phosphorite) deposits are both of igneous and sedimentary origin and they are part of the apatite group. They are commonly encountered as fluorapatite and francolite respectively. A specific characteristic of phosphate rocks consists in a low ratio Th/U, in general less than 0.5, which is mainly due to relative high concentrations of uranium commonly between 370 and 2470 Bq kg$^{-1}$ and sometimes higher than 12.35 kBq kg$^{-1}$. Uranium and thorium decay chains are generally found to be in secular equilibrium (**Menzel, 1968**).

Phosphorite is mainly processed through the so-called "wet process", which includes chemical treatments, mostly by using sulfuric acid: the products are phosphoric acid (PA) and an insoluble calcium sulfate salt called phosphogypsum (PG), with a ratio PG/PA = 5 (**Tayibi et al., 2009**). PA and PG are usually separated by filtration and reactor off-gas and vapors. These processes concentrate the trace elements in PA or PG in various amounts causing secular disequilibrium in U and Th decay chains. In PG is found approximately 80 – 90% of the $^{226}$Ra along with a high content of $^{210}$Pb and $^{210}$Po (**Carvalho, 1995**; **Beddow et al., 2006**), as a consequence of the similar chemistry (**Guimond and Hardin, 1989**; **Beddow et al., 2006**). About 80 – 85% of the uranium (**Poole et al., 1995**; **Beddow et al., 2006**) and about 70% of thorium (**Tayibi et al., 2009**) concentrate in PA. As a consequence of these processes, we expect a relevant enhancement of U concentration in PA and an abundance of Ra in PG comparable with that in phosphate-ore (**Table 3**). The high production of posphogypsum requires constant controls in order to make a secure stockage and reutilization of such material. These controls must be extended to PA and phosphate fertilizers: the MCA_Rad system described in the next section has been designed to deal with automatic gamma ray measurements on large amounts of PA and PG samples.

**Table 3**. Studies on activities of phosphate rock (PR) and phosphogypsum (PG) in different countries.

| Authors of the study | Country | $^{238}$U (Bq kg$^{-1}$) | | $^{226}$Ra (Bq kg$^{-1}$) | |
|---|---|---|---|---|---|
| | | PR | PG | PR | PG |
| **Mazilli et al., 2000** | Brazil | - | - | 130 - 1445 | 93 - 729 |
| **Silva et al., 2001** | Brazil | 434 - 1128 | 66 - 140 | 407 - 1121 | 228 - 702 |
| **Saueia et al., 2005** | Brazil | 14 - 638 | <2 - 61 | 53 - 723 | 24 - 700 |
| **Santos et al., 2006** | Brazil | 102 - 1642 | 32 - 69 | 239 - 862 | 307 - 1251 |
| **Saueia et al., 2006** | Brazil | 158 - 1868 | 40 - 58 | 139 - 1518 | 122 - 940 |
| **El Afifi et al., 2009** | Egypt | 916 | 140 | 890 | 459 |
| **Azouazi et al., 2001** | Morocco | 2100 - 2450 | - | 1850 - 2320 | 1420 |
| **Carvalho, 1995** | Portugal | 1003 | 26 - 156 | 1406 | 950 - 1043 |
| **Hull and Burnett, 1996** | USA (Central Florida) | 848 - 1980 | 45 - 368 | 882 - 1980 | 505 - 1353 |
| | USA (Northern Florida) | 242 - 982 | 23 - 452 | 230 - 883 | 270 - 598 |

## 3. MCA_Rad system

An accurate radiological characterization of NORM requires careful investigation on a case-by-case basis. Indeed the NORM issue shows high levels of complexity mainly due to the huge amount of worldwide waste production perturbed by the different geochemical composition of raw materials and the effective concentration of a wide range of industrial processes. Most of current studies show that determining the content of $^{226}$Ra, $^{228}$Ra and $^{40}$K radioisotopes in NORM is mandatory for radiation protection. High resolution gamma-ray spectrometry is a frequently used non-destructive technique, which provides an accurate identification and quantitative determination of such radioisotopes.

A strategic approach to the NORM issue consists in constructing systematic monitoring programmes. However, this approach is often limited by funding and manpower capacities of a laboratory. Indeed, when a laboratory deals with random measurements, the main costs are determined by the instrument investment programme; instead, for a routine monitoring program the substantial costs are due to the manpower involved. The MCA_Rad system introduces an innovative configuration of a laboratory high-resolution gamma-ray spectrometer featured with a complete automation measurement process. Two HPGe detectors allow to achieve both good statistical accuracy in a short time and *ad hoc* approach to low-background shielding construction design. This self-constructed instrument drastically minimizes measurement and manpower costs.

*3.1 Set-up design and automation*

The core of the MCA_Rad system is made of two 60% relative efficiency coaxial p-type HPGe gamma-ray detectors, which possess an energy resolution of about 1.9 keV at 1332.5 keV ($^{60}$Co). Both detectors are controlled by individual integrated gamma spectrometers for the digital signal processing. The new cooling technology, which employs mechanical coolers, allows to simplify the management of the system. The detectors are accurately shielded and positioned facing each other 5 cm apart (**Fig. 1a**).

The background spectrum of a gamma ray spectrometer is mainly due to the combination of cosmic radiation, environmental gamma radiation and the radioactivity produced by radio-impurities both in the shielding materials and in the detector. In order to effectively reduce the environmental gamma radiation, an adequate shielding construction is needed.

In the MCA_Rad a 10 cm thick lead house shields the detector assembly, leaving an inner volume around the detectors of about 10 dm$^3$ (**Fig.1a** and **1b**). The lead used as shielding material adds some extra background due to the presence of $^{210}$Pb, produced by $^{238}$U decay chain. This isotope, which has an half life of 22.3 years, is revealed by a gamma energy of 46.5 keV and a *bremsstrahlung continuum* from beta decay of its daughter $^{210}$Bi extending from low energy up to 1162 keV. Furthermore, when a gamma ray strikes the lead surface, characteristic lead X-rays may escape and hit the detector (**Smith et al., 2008**).

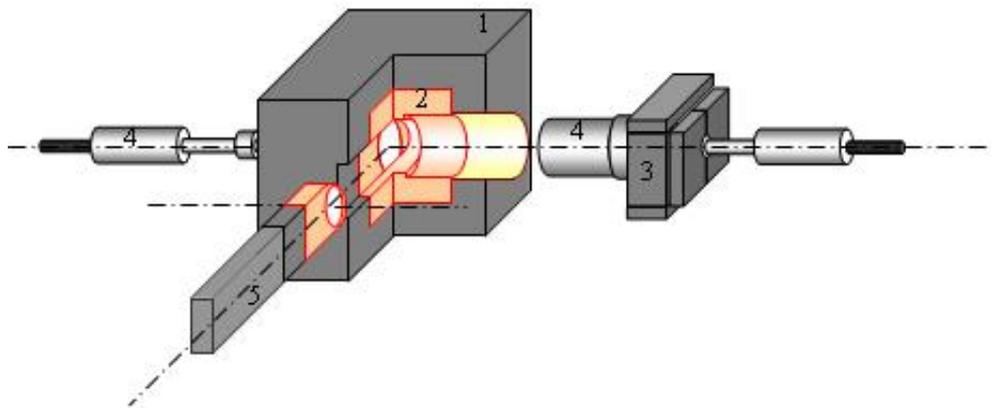

**Figure 1a.** Schematic design of the MCA_Rad system. 1) The main lead shielding construction (20 cm x 25 cm x 20 cm). 2) The core copper shielding (10 cm x 15 cm x 10 cm). 3) Rear lead shielding construction. 4) HPGe semiconductor detectors. 5) The mechanical sample changer.

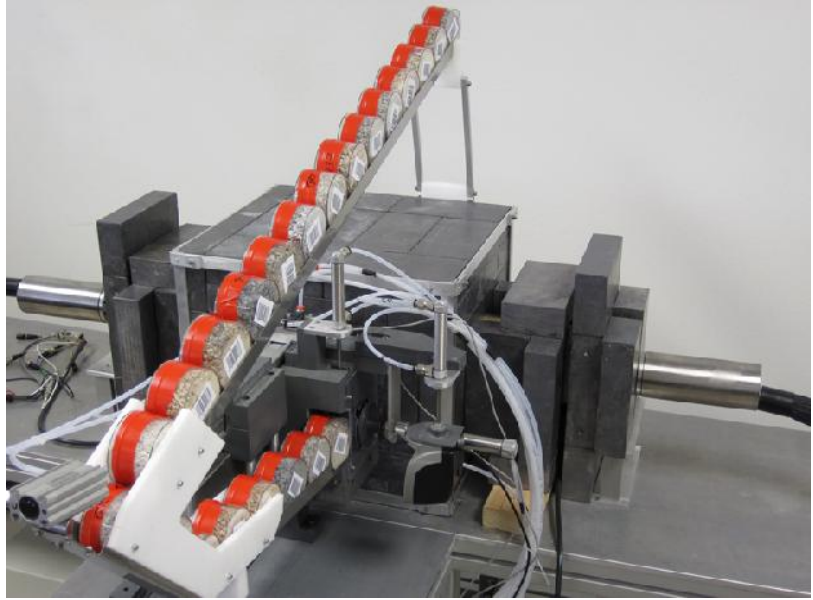

**Figure 1b.** View of the MCA_Rad system.

The inner volume is occupied by 10 cm thick oxygen free copper house, which allows to host the sample under investigation. In order to reduce the X-rays coming from the sample, the end-cup windows of the detectors is further shielded with a tungsten alloy sheet of 0.6 mm. A 10 mm thick bronze cylinders and walls of about 10 cm of lead are also shielding the rear part of the system. **Fig.1a** and **1b**). The final intrinsic background is reduced by two order of magnitude compared to other unshielded detectors (see **Fig. 2**).

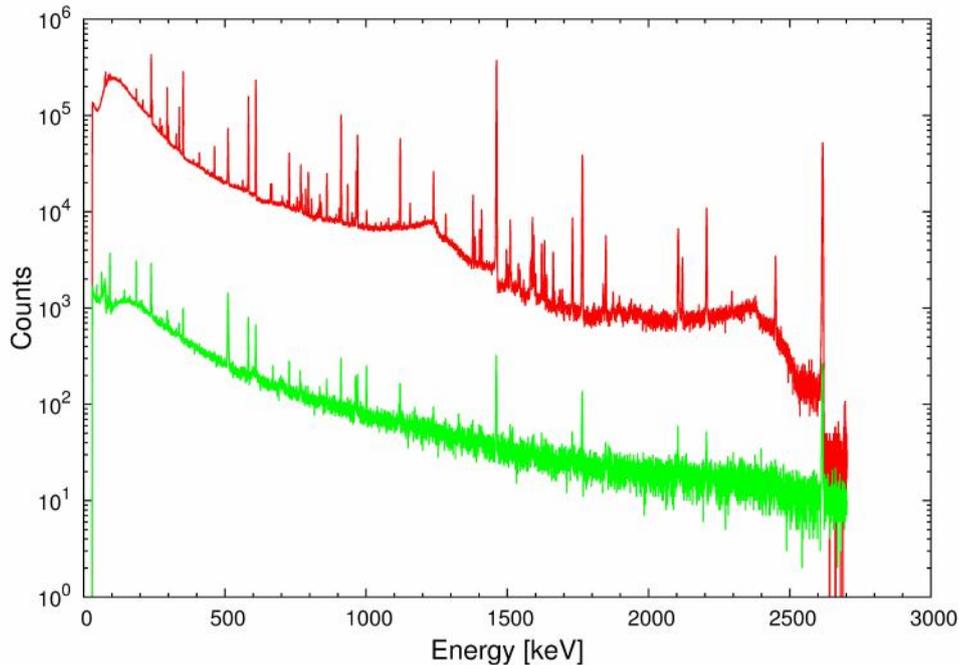

**Figure 2**. The MCA_Rad system background spectra (acquisition live time 100 h) with (green) and without (red) shielding. Spectra are obtained by summing the single detector background after rebinning with 0.33 keV/channel.

As a good practice of gamma-ray spectrometry analysis, information concerning background spectra is required both to detect any potential residual contaminations and for background corrections. A background measurement with acquisition time of several days is performed regularly. The final sensitivity of the measurements can be evaluated by using the detection limit ($L_D$) described in (**Currie, 1986**), assuming the Gaussian probability distribution of the number of counts in the background (B) and rejecting the data not included in a range of 1.645σ (95% confidence level):

$$L_D = 2.71 + 4.65\sqrt{B} \approx 4.65\sqrt{B} \tag{1}$$

where the approximation is admitted for high number of counts. The minimum detectable activity (MDA) for the background is calculated using the $L_D$, according to the formula:

$$MDA = \frac{L_D}{\varepsilon I_\gamma t} \tag{2}$$

where $\varepsilon$ is the absolute efficiency (calculated as described below), $I_\gamma$ is the gamma line intensity and $t$ is the acquisition live time. In **Table 4** we report the typical one hour acquisition live time background counts and the sensitivity of the measurement expressed by $L_D$ and MDA for the main gamma lines used to calculate the radionuclide concentrations in NORM.

**Table 4.** MCA_Rad system characterization of typical one hour (live time) background (B in counts) for the most important energies and the corresponding detection limit $L_D$ and minimum detectable activity (MDA) (**Currie, 1986)** for 95% confidence interval (CI).

| Parent isotope | Daughter isotope | Energy (keV) | B (counts) | $L_D$ (counts) | MDA (Bq) |
|---|---|---|---|---|---|
| $^{238}$U | $^{234m}$Pa | 1001.0 | 8 ± 1 | 21 | 22.16 |
| | $^{214}$Pb | 351.9 | 31 ± 2 | 49 | 0.50 |
| | $^{214}$Bi | 609.3 | 44 ± 1 | 32 | 0.49 |
| $^{232}$Th | $^{228}$Ac | 911.2 | 27 ± 1 | 27 | 0.94 |
| | $^{212}$Pb | 238.6 | 100 ± 2 | 62 | 0.46 |
| | $^{212}$Bi | 727.3 | 10 ± 1 | 31 | 3.00 |
| | $^{208}$Tl | 583.2 | 42 ± 1 | 33 | 0.71 |
| $^{40}$K | $^{40}$K | 1460.8 | 151 ± 1 | 19 | 5.53 |

The sample material is contained in a cylindrical polycarbonate box of 75 mm in diameter, 45 mm in height and 180 cm$^3$ of useful volume, labeled by a barcode. Up to 24 samples can be charged in a slider moving by gravity and further introduced at the inner chamber through an automatic "arm" made of copper, lead and plastic closing the lateral hole of the housing (**Fig. 1b**). The mechanical automation consists on a barcode scanner and a set of compressed air driven pistons. This mechanism not only makes the sample identification possible, but is also able to introduce/expel the samples. All operations, including measurements, are controlled by a PC by means of a dedicated software.

The program receives by the operator an input file with relevant information about the slot of samples: acquisition live time, spectra file name, sample weight, sample description and barcode. The procedure is repeated until the barcode reader detects samples. A new batch command file is generated to be successively employed in spectrum analysis. In order to complete the automation of the MCA_Rad system, a user-friendly software has been developed for spectra analysis. The code adopts **ANSI No. 42.14, 1999** standard specification for the peak analysis.

*3.2 Calibration and data analysis*

For each measurement, the final spectrum is obtained by adding, after rebinning, the two simultaneously measured spectra: for this purpose an accurate energetic calibration of the system, along with a periodical check, is required. When a shift larger than 0.5 keV is observed, the energy calibration procedure is repeated.

The absolute photopeak efficiency ($\varepsilon_P$) for the MCA_Rad system has been determined by using standard point sources method, and producing the calibration curve. Two low activity point sources with complex decay schemes are used (**DeFelice et al., 2006**): a certified $^{152}$Eu source, with an activity of 6.56 kBq in 1995, known with an uncertainty of 1.5% and a $^{56}$Co home made source, that has been normalized relative to $^{152}$Eu by calculating the activity of the 846.8 keV ($^{56}$Co) gamma line. The $^{56}$Co source is used in order to extend the efficiency calibration for gamma energies up to 3000 keV.

The spectra obtained are corrected for: 1) coincidence summing, $C_{CS}$, on each individual detector, produced by the complex decay scheme of the sources, 2) differences between the geometry of the point sources and the sample shape, $C_G$ and 3) self-attenuation, $C_A$, of gamma-rays within the sample volume.

The correction due to coincidence summing is studied by following the method described in (**DeFelice et al., 2000**) and obtained as a relationship between the single total efficiency ($\varepsilon_t$), the single apparent absolute photopeak efficiency ($\varepsilon_p^{app}$) and isotope decay data. The single total efficiency is obtained by estimating the peak-to-total ratio ($P/T$) using the empirical approach described by (**Cesana and Terrani, 1989**) and recalling the relationship $\varepsilon_p^{app}/\varepsilon_t = P/T$. The decay coefficients for $^{152}$Eu were calculated from decay data taken from **Monographie BIPM-5, 2004**, while those for $^{56}$Co were taken from **Tomarchio and Rizzo, 2011** and **Dryák and Kovář, 2009**. Finally, the absolute efficiency, $\varepsilon_p(E)$, of the MCA_Rad system is given by the sum of single HPGe detector contribution corresponding to the characteristic gamma energies ($E_i$) of the standard calibration sources used.

$$\varepsilon_p(E_i) = \varepsilon_p^{app}(E_i) C_{CS}(E_i) \qquad (3)$$

Then the absolute efficiency is obtained for the energetic range from 200 keV to 3000 keV by fitting them using the function given by **Knoll, 1999** (**Fig. 3**):

$$\varepsilon(\%) = \left(b_0 E/E_0\right)^{b_1} + b_2 \exp\left(-b_3 E/E_0\right) + b_4 \exp\left(-b_5 E/E_0\right) \qquad (4)$$

where E (keV) is the gamma-ray energy; $E_0$ = 1 keV is introduced to make dimensionless the argument of the exponential dimensionless and $b_i$ are the fitting parameters (where $b_0$ = 1.38, $b_1$ = 1.41, $b_2$ = 22.97, $b_3$ = 5.43, $b_4$ = 6.61 and $b_5$ = 0.44).

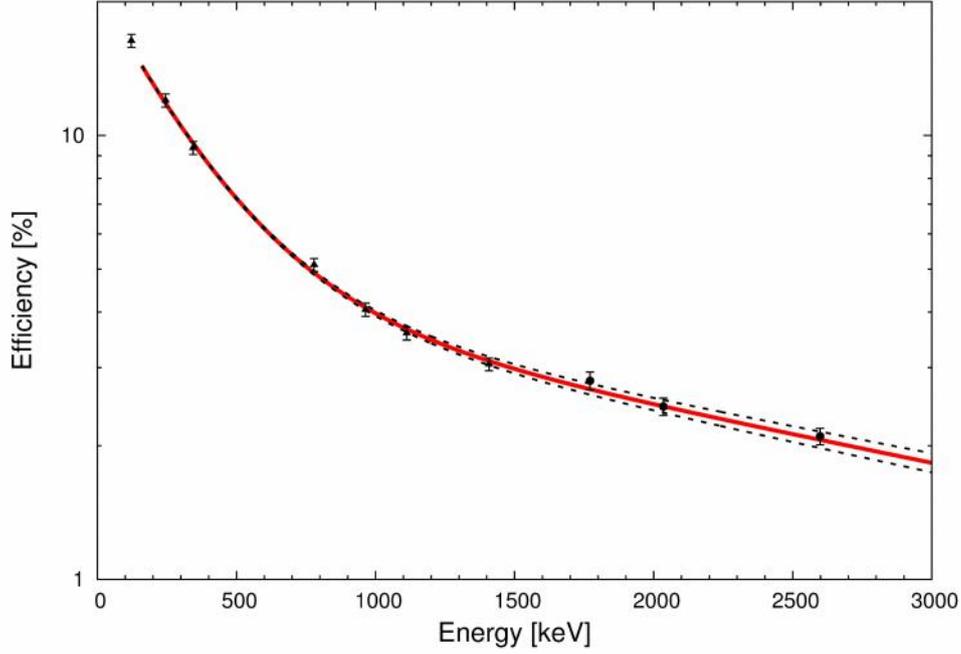

**Figure 3.** Absolute efficiency curve of MCA_Rad system. It is obtained by fitting $^{152}$Eu (triangle) and $^{56}$Co (circle) energies with Eq. 2 performing the best fit (in red). Dashed curves represent ± one sigma uncertainty interval.

The effect of volume geometry can be described in terms of the effective solid angle developed analytically by **Moens et al., 1981** within less than 2% of uncertainty between numerical and experimental calculations. The geometrical factor (**Fig. 4**) for MCA_Rad system is deduced from a set of measurements using a $^{56}$Co and $^{57}$Co point sources placed at different radial distances (center, middle and lateral) from the detector axis at different planes, reconstructing the sample geometry using the formula:

$$C_G(E_i) = \frac{\overline{\Omega}_x}{\overline{\Omega}_{ref}} \approx \frac{1}{R_{ref}(E_i)} \sum_{j=1}^{N} \frac{[R_x(E_i)]_j}{N} \quad (5)$$

where $R_x(E_i)$ is the net count rate in the standard spectrum collected in different positions (j) and $R_{ref}(E_i)$ is the net count rate in the standard spectrum collected in the reference positions (center). The geometrical correction factor is obtained as a function of energy by fitting a third order polynomial as a function of energy (**Fig. 4**):

$$C_G(E) = \sum_{i=0}^{4} a_i \left(\frac{E}{E_0}\right)^i \quad (6)$$

where $E_0$ = 1 keV is introduced to make the argument dimensionless and $a_i$ are the fitting coefficients ($a_0$ = 0.8678, $a_1$ = 0.1098, $a_2$ = -0.0541, $a_3$ = 0.0077).

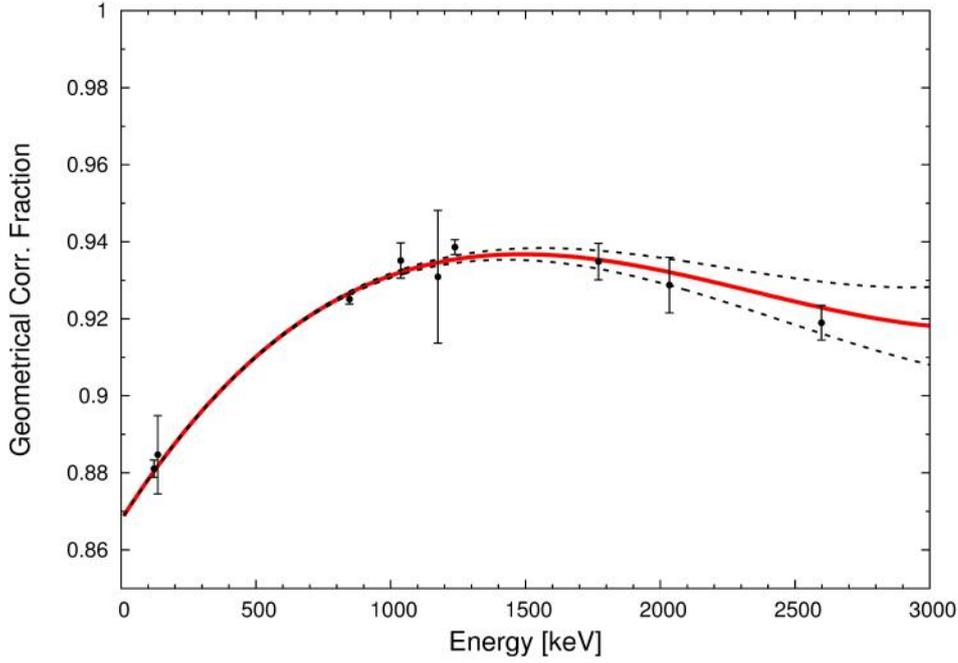

**Figure 4.** The geometrical correction factor due to differences between calibration (point source) and measurement geometry (cylinder of 75 mm of diameter and 45 mm of height), which is obtained by fitting nine photopeaks of $^{56}$Co and $^{57}$Co with a third order polynomial. Dashed curves represent ± one sigma uncertainty interval.

The gamma-ray attenuation correction $C_{SA}$ is calculated for different kind of samples taking into account their differences in density by a simplified expression deduced by **Cutshall et al, 1983** and **Bolivar et al, 1997**:

$$C_{SA} = \frac{1 - \exp(-\mu_l t)}{\mu_l t} \qquad (7)$$

where $\mu_l = \mu\rho$ (cm$^{-1}$) is the linear mass attenuation coefficient, $\mu$ (cm$^2$ g$^{-1}$) is the mass attenuation coefficient, $\rho$ (g cm$^{-3}$) is the sample density and t (cm) is the sample effective thickness (which in our case is the half thickness of the sample container). The mass attenuation coefficient is strongly Z dependent in the energy range below few hundred keV while for higher energies the trend is smoother and it depends mainly on energy. These features can be used, since NORM characterization, especially concerning $^{226}$Ra, $^{228}$Ra and $^{40}$K, requires the investigation of gamma-rays with energies higher than hundreds keV. We can parameterize the mass attenuation coefficient as a function of energy. We used XCOM 3.1 database, which is available on-line and developed by the Nuclear Institute of Standards and Technology (NIST). It is calculated by using For various rocks forming minerals, the average mass attenuation coefficient was deduced with a standard deviation of less than 2% in the energetic range 200 – 3000 keV.

$$\bar{\mu}(E) = \sum_{i=0}^{2} a_i [\ln(E)]^i \qquad (8)$$

where $a_i$ ($a_0$ = 0.5593, $a_1$ = -0.1128, $a_2$ = 0.0590) are coefficients determined by fitting this function with a reduced chi-square of $\chi_\nu^2 = 1.12$. Finally the self attenuation correction factor was given as a function of gamma-ray energy and sample density through the relationship:

$$C_{SA}(E, \rho_s) = \exp\left[b_0 + b_1 \ln(E) + b_2 \ln(E)^2\right] \rho_s \quad (9)$$

where $b_i$ ($b_0$ = 1.2609, $b_1$ = -0.2547, $b_2$ = 0.0134) are coefficients determined by fitting this function.

The uncertainty budged for the calibration procedure is reported in **table 5**. Considering the uncertainties due to geometrical and self-attenuation correction factors as systematic errors, the overall uncertainty about the absolute efficiency of the MCA_Rad system is estimated to be less than 5%.

**Table 5**. Uncertainty budged for absolute efficiency determination.

| Uncertainty source | Relative uncertainty (%) |
| --- | --- |
| Certified standard source uncertainty | 1.5* |
| Coincidence summing correction factor | < 2 |
| Geometrical correction factor | < 2 |
| Self attenuation correction factor | < 2 |

* 95% of confidence level.

### 3.3 Measurement of reference materials

The applicability of the MCA_Rad system as well as the method quality control was cross checked using certified reference materials containing concentrations comparable to NORM values. Three reference materials certified within 95% of confidence level prepared in powder matrix (240 mesh) containing $^{238}$U (IAEA RGU-1) 4940 ± 30 Bq kg$^{-1}$, $^{232}$Th (IAEA RGTh-1) 3250 ± 90 Bq kg$^{-1}$ in secular equilibrium and $^{40}$K (IAEA RGK-1) 14000 ± 400 Bq kg$^{-1}$ are used. The sample boxes were filled with the reference materials, after were dried at 60°C temperature, hermetically sealed and then left undisturbed for at least 3 weeks in order to establish radioactive equilibrium in $^{226}$Ra decay chain segment prior to be measured.

In **Table 6** we report the specific activity calculated for the principal gamma lines used to estimate the isotopes of uranium and thorium decay chain and for potassium using the formula:

$$A(Bq/kg) = \frac{R}{\varepsilon I_\gamma m} C_{SA} C_G C_{CS}^* \quad (10)$$

where $R$ is the measured count rate (background corrected), $\varepsilon$ is the absolute efficiency, $I_\gamma$ is the gamma line intensity, $m$ is the mass of the sample, $C_{SA}$ is the correction factors for self-absorption, $C_G$ is the geometrical correction factor and $C_{CS}^*$ is the coincidence summing correction factor (calculated using the same approach as described above for the specific decay chains of the uranium and thorium). The results have an overall relative discrepancy of less than 5% among certified central values within the reported uncertainty.

**Table 6**. In the sixth column we report the activity concentrations (in Bq kg$^{-1}$) calculated for the main energetic lines used for $^{238}$U and $^{232}$Th decay chains and for $^{40}$K together with respective statistical uncertainties. The reference material activities certificated by IAEA are shown in seventh column. The correction coefficients $C_{CS}^*$ and $C_{SA}$ are referred to coincidence summing and self absorption respectively.

| Parent Isotope | Daughter Isotope | E (keV) | $C_{CS}^{*}$ | $C_{SA}$ | Activity (Bq kg$^{-1}$) | Certified Reference Material Activity (Bq kg$^{-1}$) |
|---|---|---|---|---|---|---|
| $^{238}$U | $^{234m}$Pa | 1001.0 | 1.000 | 1.24 | 4875 ± 48 | 4940 ± 30 |
|  | $^{214}$Bi | 609.3 | 1.190 | 1.32 | 4872 ± 4 |  |
|  | $^{214}$Pb | 351.9 | 1.002 | 1.42 | 4773 ± 3 |  |
| $^{232}$Th | $^{228}$Ac | 911.2 | 1.024 | 1.24 | 3092 ± 4 | 3250 ± 90 |
|  | $^{212}$Pb | 238.6 | 0.990 | 1.48 | 3246 ± 2 |  |
|  | $^{212}$Bi | 727.3 | 1.056 | 1.27 | 3389 ± 9 |  |
|  | $^{208}$Tl | 583.2 | 1.298 | 1.31 | 3342 ± 4 |  |
| $^{40}$K | - | 1460.8 | - | 1.21 | 14274 ± 71 | 14000 ± 400 |

## 4. Conclusions

We presented a summary of the main categories of non-nuclear industries together with the levels of activity concentration in feeding raw materials, products, by-products/waste and the possible enhancement mechanisms. The main chemical and physical processes that disturb the secular equilibrium of uranium and thorium decay chains have been reported. The degree of radioactivity enhancement is studied along with the radioactivity level in feeding raw material and the industrial process involved. A refined estimation of radioactivity concentrations of $^{226}$Ra, $^{228}$Ra and $^{40}$K in NORM is almost impossible in such a wide range of industrial activities: the strategy that we propose consist in a systematic monitoring and continuous checking based on high-resolution gamma-ray spectroscopy.

For this purpose an innovative approach to the configuration of a laboratory low background high-resolution gamma-ray spectrometer, MCA_Rad system, was developed and featured with fully automated measurement processes. It presents the following advantages.

- The new design of lead and cooper shielding configuration allowed to reach a background reduction of two order of magnitude respect to laboratory radioactivity.
- A severe lowering of manpower cost is obtained by a fully automation system which permits to measure up to 24 samples without any human attendance.
- The two HPGe detectors offer higher detection efficiency: confronting the MDA of the system with typical NORM values, it can be deduced that measurements in less than one hour are realized with uncertainties of less than 5%.
- Accurate measurements are performed on small sample volume (180 cc) with a reduction of material transport costs.
- A user-friendly software has been developed in order to analyze a high number of spectra, possibly with automatic procedure and customized output.

An empirical efficiency calibration method using multi-gamma standard point sources is discussed. The correction factors affecting the measured spectra (coincidence summing, sample shape, sample gamma-ray self-attenuation) are given with respective procedures. As a result of this procedure the absolute efficiency is estimated to have an overall uncertainty of less than 5%. A test of the applicability of the instrument as well as the method quality control using certified reference materials showed an overall relative discrepancy of less than 5% among certified central values within the reported uncertainty.

We, therefore, conclude that the MCA_Rad system shows efficacy in the face of the NORM issue, by increasing the capacities of a laboratory and offering accurate results with a reduction of manpower costs.

**Acknowledgement**

We are grateful for useful comments and discussions to P. Altair, M. Baldoncini, E. Bellotti, L. Carmignani, L. Casini, V. Chubakov, T. Colonna, M. Gambaccini, Y. Huan, W. F. McDonough, G. Oggiano, R. L. Rudnick and R. Vannucci.

This work was partially supported by INFN (Italy) and by Fondazione Cassa di Risparmio di Padova e Rovigo.

WCI World Coal Institute, Coal Facts 2005 Edition.